# Feasibility for $p^+/p^-$ flow-ratio evaluation in the 0.5 – 1.5 TeV primary energy range, based on Moon-shadow muon measurements, to be carried out in the Pyramid of the Sun, Teotihuacan, Experiment

V Grabski[1], A Morales[1], R Reche[2], O Orozco[1]
[1] *Instituto de Física, Universidad Nacional Autónoma de México, Mexico*
[2] *Laboratoire de Physique Subatomique et de Cosmologie, Grenoble, France*
*varleni.grabskil@cern.ch*

## Abstract

Calculations are presented to demonstrate the feasibility of Moon shadow observations for mean primary energies in the region 0.5-1.5 TeV using a muon detector operating under the Pyramid of the Sun at Teotihuacan, Mexico. Due to the small height of that monument (65 m), the experiment is capable of providing considerably high statistics, although with reduced angular accuracy for primary particle direction reconstruction. Our estimates are based on simulations of muon production and transport in the atmosphere by CORSIKA and along the body of the pyramid by GEANT4. The deflection of primaries in the earth magnetic field is calculated using the IGRF model. The statistics for the Moon shadow observations, which depends on different factors affecting the accuracy of the primary particle direction reconstruction, are analyzed in detail.

## Introduction

The antiproton-to-proton ratio ($p^-/p^+$) in primary cosmic rays is relevant not only to understand the acceleration and propagation of cosmic rays in the interstellar space but also to search for possible antimatter objects in the universe. Other exotic sources of high energy antiprotons are: primordial black hole evaporation [1], dark-mater neutralino annihilation [2], and high-energy antiprotons from extragalactic sources [3].

In the low-energy region $\leq$ 10 GeV direct observations have been performed using balloon-born detectors [4, 5]. In the energy region > 10 GeV, the $p^-/p^+$ ratio was indirectly estimated to be $\leq$ 5% from the observed $\pi/p$ ratio [6]. From the observed muon charge ratio, the $p^-/p^+$ ratio was also indirectly estimated to be 17% at energies $\approx$ 1 TeV, 10% at 10-15 TeV and 14% at energies > 30 TeV [3].

The first direct $p^-/p^+$ ratio estimation using Sun-shadowing was reported in 1995 by the Tibet AS$\gamma$ collaboration for the energy region $\approx$ 10 TeV and found to be $\leq$ 28% [7-9]. For the energies $\approx$ 1 TeV the first direct $p^-/p^+$ preliminary estimate was reported in 2002 by the Tibet AS$\gamma$ [10] and L3C collaborations [11,12]. There are no other results between 30 GeV and 1TeV, where this ratio is expected to increases two orders of magnitude.

The archaeological experiment at the Pyramid of the Sun in México [13] will collect more than $10^8$ underground muons with energies $\geq$ 20 GeV. We have studied the mean energy region of primaries 0.7-1.5 TeV by means of Monte Carlo simulations to define the accuracy in determining primary directions from secondary muons. In this region the determination of the primary particle direction is less accurate but provides better statistics. Using the obtained accuracies, the expected signal to noise ratios (SNR) for detection the Moon shadow using simulation have been estimated.

## Method

There are two main factors affecting the detection of the Moon-shadow signal: the angular spread due to particle deflection for primaries having different momentum, and the angular spread during the production and transport of the muons. All the above-mentioned factors depend on the muon minimal energy. The pyramid of the Sun has a complicated external shape and the detector is located 40 m away from the geometrical center, so the minimal energy of the muons penetrating into the body of the pyramid depends on their direction. The Monte Carlo simulation has been performed using the external shape of the pyramid, and assuming an homogeneous internal density distribution. The angular dependence of the muon minimal energy for the Moon position as a function of the Y-projected angle ($\theta_y$) is presented in Fig 1. The projection angles were calculated along (X) and perpendicular (Y) to the direction of the geomagnetic field.

The deflection-angle for the primary particle energy > 1TeV was previously calculated [9]. For the vertical direction $\Theta_{def}= 1.6°/E(TeV)$. We have repeated this calculation using IGRF model for Mexico City coordinates. The obtained results for the deflection angle dependence on the energy and the zenith angle of the primary protons can be expressed with the polynomial $(\Theta_{def}(E, \theta_y) = 1471.075-0.83362\theta_y +0.18022\theta_y^2)/E(GeV)$. The deflection angular distribution of primaries for the given muon minimal energy can be estimated using the energy distribution of primaries, based on CORSIKA simulations. The primaries were generated in the energy interval 20-$10^6$ GeV for the zenith angle interval 0-70°. In simulations three hadronic models are considered (QGSJET, VENUS and HPDM). The simulation data have been analyzed up to a muon minimal energy of 500 GeV. The distribution of the deflection angle can be well described using an exponential function exp(ax+b) (see Fig 2). The slope of the exponent ***a*** linearly depends on the muon minimal energy ($a=0.007E_\mu^{min}+0.022$ [deg$^{-1}$], where $E_\mu^{min}$ is expressed in GeV).

To study the angular spread between the primary proton and the secondary muon during the production, as well as the transport up to earth surface, the distributions of projection angles have been constructed. The distributions can be described well enough by the sum of two Gaussian (see Fig 3). The resulting standard deviations of the narrow Gaussian ($\sigma^n_{\theta x}$) and the wide Gaussian ($\sigma^w_{\theta y}$) of X projection angle dependent on the muon minimum energy is presented in Fig 4 with the relative weights. As can be seen from the figure, the results of different models coincide within 20 %, and have approximately the same energy dependence on the muon minimal energy $\sigma^n_{\theta x}(E_\mu^{min}) = 3.4E^{-0.78}$ and $\sigma^w_{\theta x}(E_\mu^{min}) = 12.6E^{-0.83}$. Our result is: for $\sigma^n_{\theta x}(E_\mu^{min} =100GeV)=0.094°$ and $\sigma^w_{\theta x}(E_\mu^{min} =100GeV)=0.28°$, while the experimental result of the L3C [12] collaboration for total angular resolution is 0.22°.

Total event deficit signal from the Moon shadow will be blurred by the distributions of the above-mentioned process. Taking into account that the signal spread area is proportional to the standard deviations of the angles in both directions (parameter a has stronger dependence on the muon minimal energy than $\sigma^w_{\theta y}$ so it dominates) one can roughly estimate the signal to noise ratio SNR dependence on the muon minimal energy:

$$SNR(E_\mu^{min}) \sim E^{(2.78-\gamma)/2}$$

Where $\gamma$ is the slope of the differential muon intensity. This clearly shows that SNR has very weak dependence on muon minimal energy.

To determine the total angular spread in underground measurements it is necessary to determine the one corresponding to the underground layer conditioned by multiple scattering. This effect was estimated by use of GEANT4 simulation. The muon multiple scattering was studied for different thickness of soil matter and for different energy cuts in underground. The simulation results for the standard deviation (single Gaussian is not appropriate) of the distribution of the projected angles depending on the layer thickness and for different muon energy cuts in underground is presented in Fig. 5.

Having all components of the spread in the primary particle direction, the antiproton deficit signal extraction can now be discussed. From deflection angle distribution it is clear that the antiproton should be estimated as an excess on the positive particle shadow deficit signal. Taking into account that the expected antiproton signal is $\sim$ 10% [12] from the proton signal, the required statistics for the antiproton shadow detection for the same confidence level should be at least 100 times more. The total signal parameterization and their determination by minimization methods [12] can be accurate if the exact distributions are known. In our estimations the *1/a* is always larger than $3\sigma^{w}$ and if the detector resolution is sufficiently good the spatial separation of the antiproton signal is possible. But in this case it is necessary to have significantly more statistics.

In underground experiments it is possible to have good angular resolution if there are possibilities to use large energy cuts. In the Teotihuacan experiment we expected to have a total detector resolution, with multiple scattering, better than 0.6°, which is still smaller than *1/a* for the energy interval $E_{\mu}^{min}$ 20-50 GeV.

## Results

We have simulated the Moon shadow using fast simulation method based on the usage of the above-mentioned angular distributions. At the beginning for the validation purpose we have repeated the simulation using L3C conditions at $E_{\mu}^{min} = 100GeV$ [12]. Using our parameterizations for a muon minimal energy of 100 GeV the obtained result for SNR is 20% lower than the L3C experimental data [12] as we anticipated taking into account that our distributions of the angular spread are wider than it was obtained in the experiment L3C [12].

In the Teotihuacan experiment the muon minimal energy varies depending on the Moon position within 20 - 50 GeV, and there is no possibility to compensate this using large energy cuts in the underground. The simulation result for the event deficit SNR depending on $\theta_y$-$\theta_y^{moon}$ angle is presented in Fig 6 for the large amount of statistics, to demonstrate the model independent observation of the shadow from the antiprotons assuming that $p^-/p^+$ is of order 0.11 [12]. The left and right sides of the shadow deficit signal starting $3\sigma_y$ from the zero angle should have the same angular dependence that can be used for the antiproton signal extraction. This is less sensitive to the shadow signal presentation but requires higher statistics.

The results of Moon shadow signal collection time for different muon minimum energy values with the corresponding SNR values for the pyramid experiment is presented in following Table. The pyramid detector has 1 $m^2$ surface and a zenith angle acceptance of ~56°. For the collection time estimation we have used parameterization [14] for muon differential intensity corrected for the Mexico City altitude.

## Conclusions

Using CORSIKA simulations the accuracy of reconstruction primary particle direction by secondary muon up to energies 500 GeV has been studied. It has been shown that the muon low minimum energy underground experiments also can be useful for antiproton flux estimation by use of the Moon shadow.

## Acknowledgements

Authors acknowledge to Dr. A. Menchaca-Rocha for the helpful discussion and the partial support from PAPIIT-UNAM Grant IN115107.

**Table** The collection time for the different muon minimum energy values.

| $\Delta E_\mu^{min}$ (GeV) | $E_\mu^{mean}$(GeV) | $\Delta\theta_y$° | SNR | Time(y) |
|---|---|---|---|---|
| 20-50 | 27 | 113 | 5 | 2 |
| 20-25 | 22 | 45 | 5 | 5. 2 |

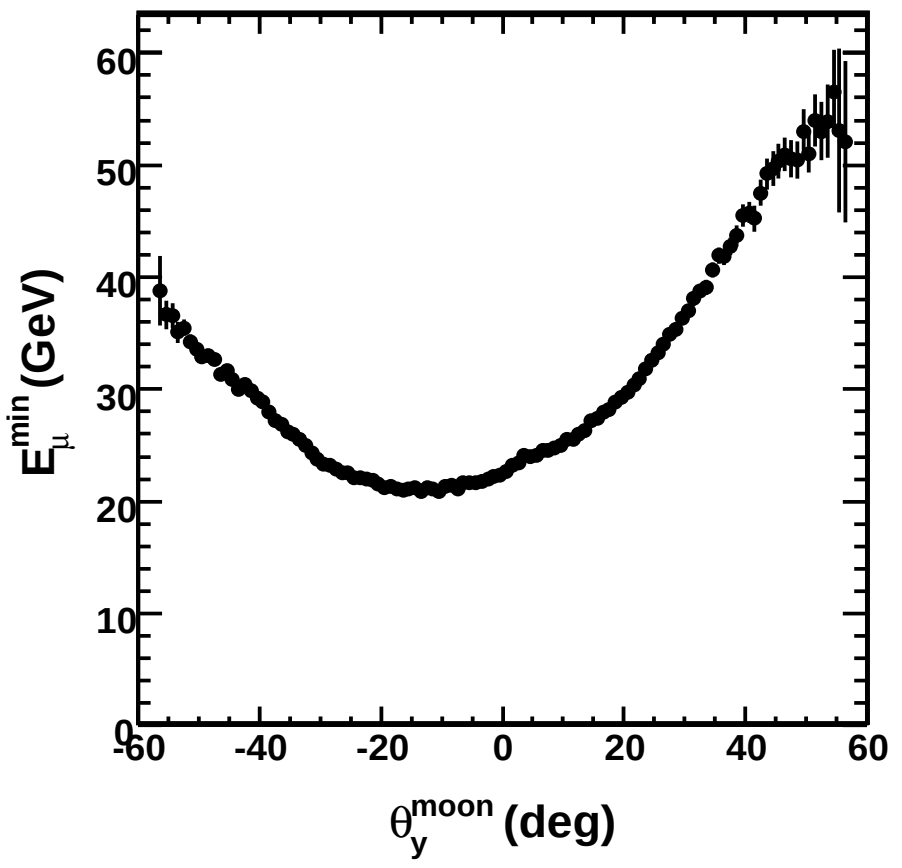

**Fig 1** The dependence of muon minimum energy on $\theta_y$ projection angle

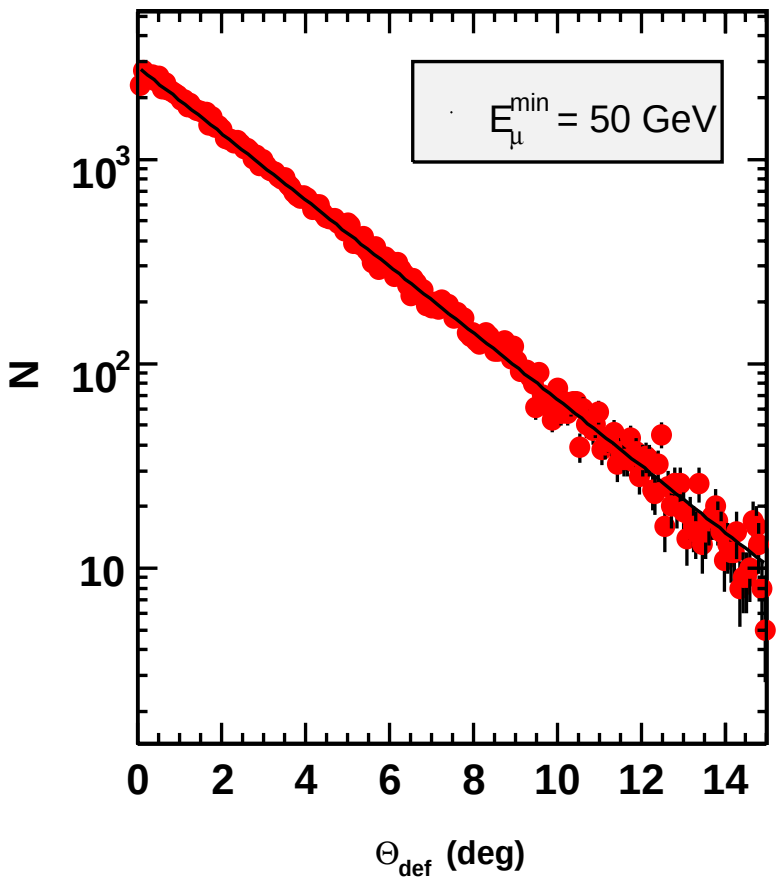

**Fig 2** Deflection angle distribution for a muon minimum energy of 50

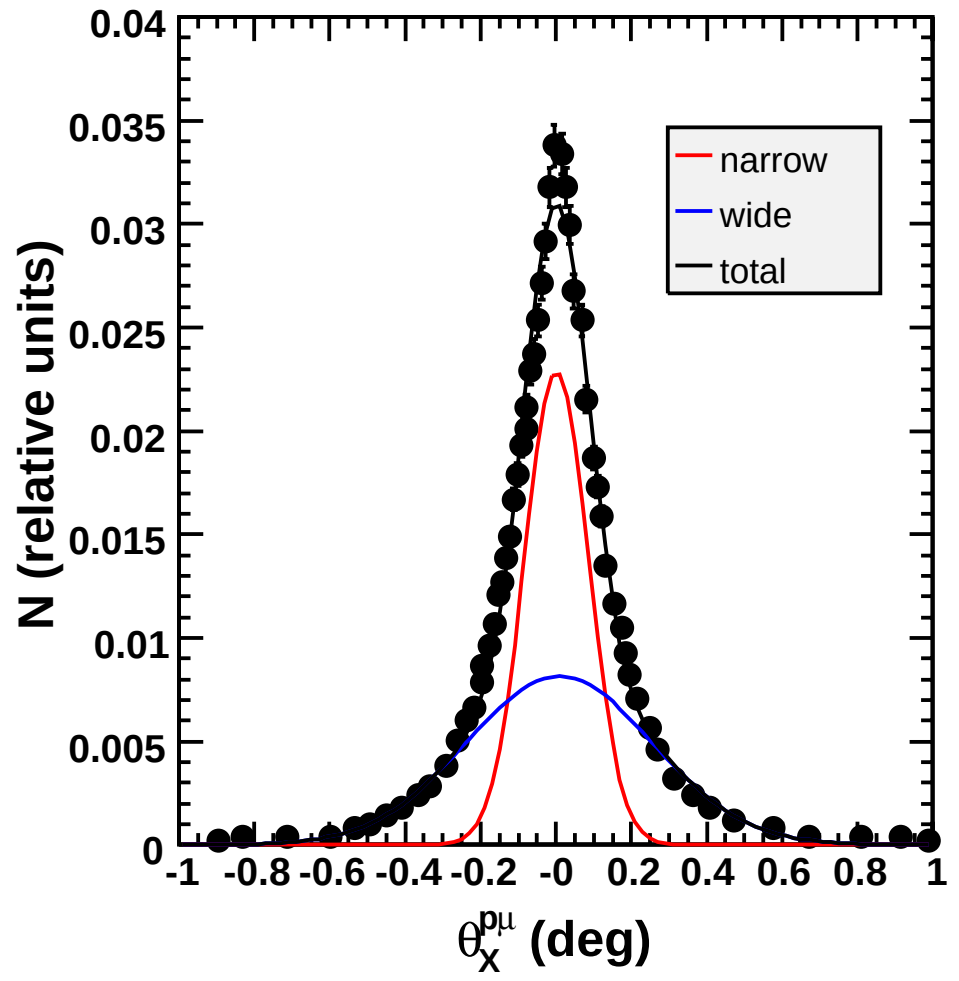

**Fig 3** X projection angle distribution for a muon minimum energy of 50 GeV.

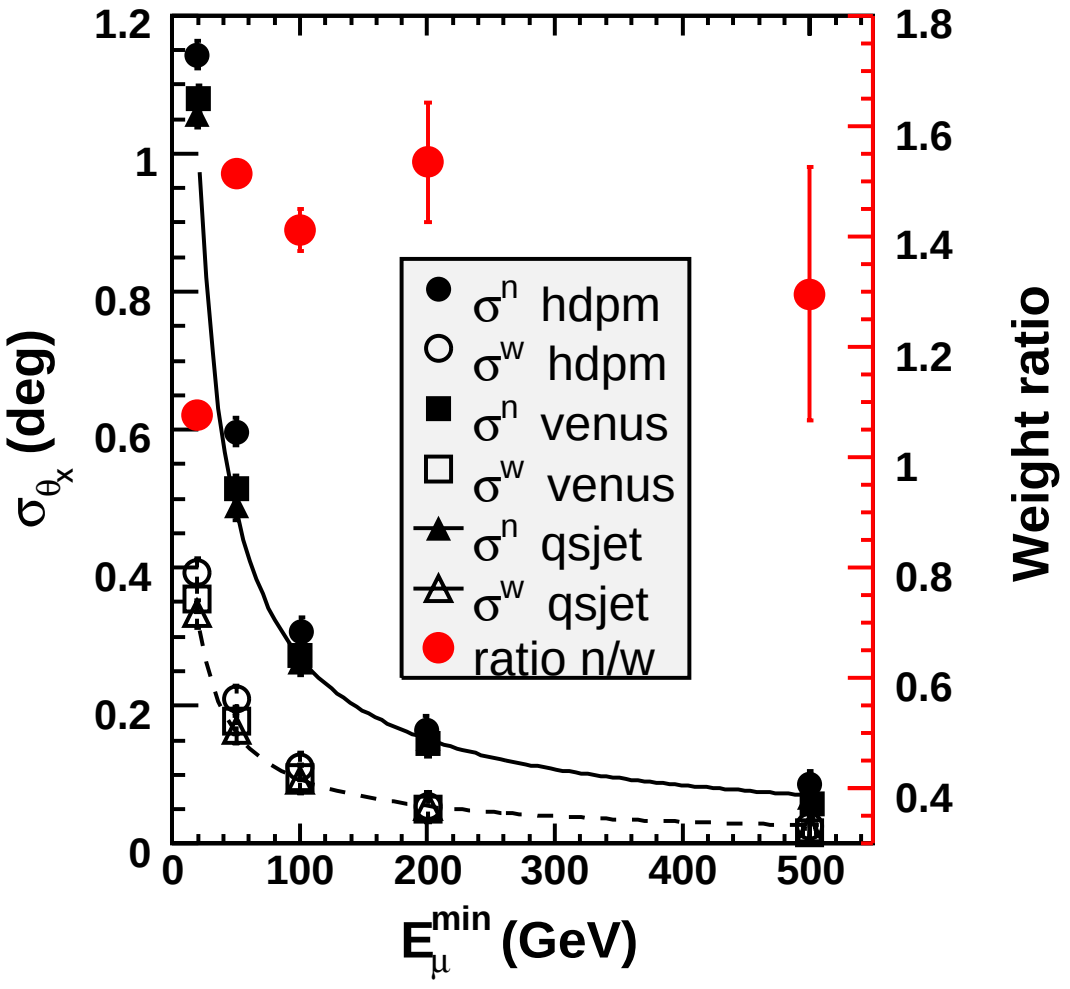

**Fig 4** $\sigma^n{}_{\theta x}$ and $\sigma^w{}_{\theta x}$ dependence on the muon minimum energy with the relative weights. (see text).

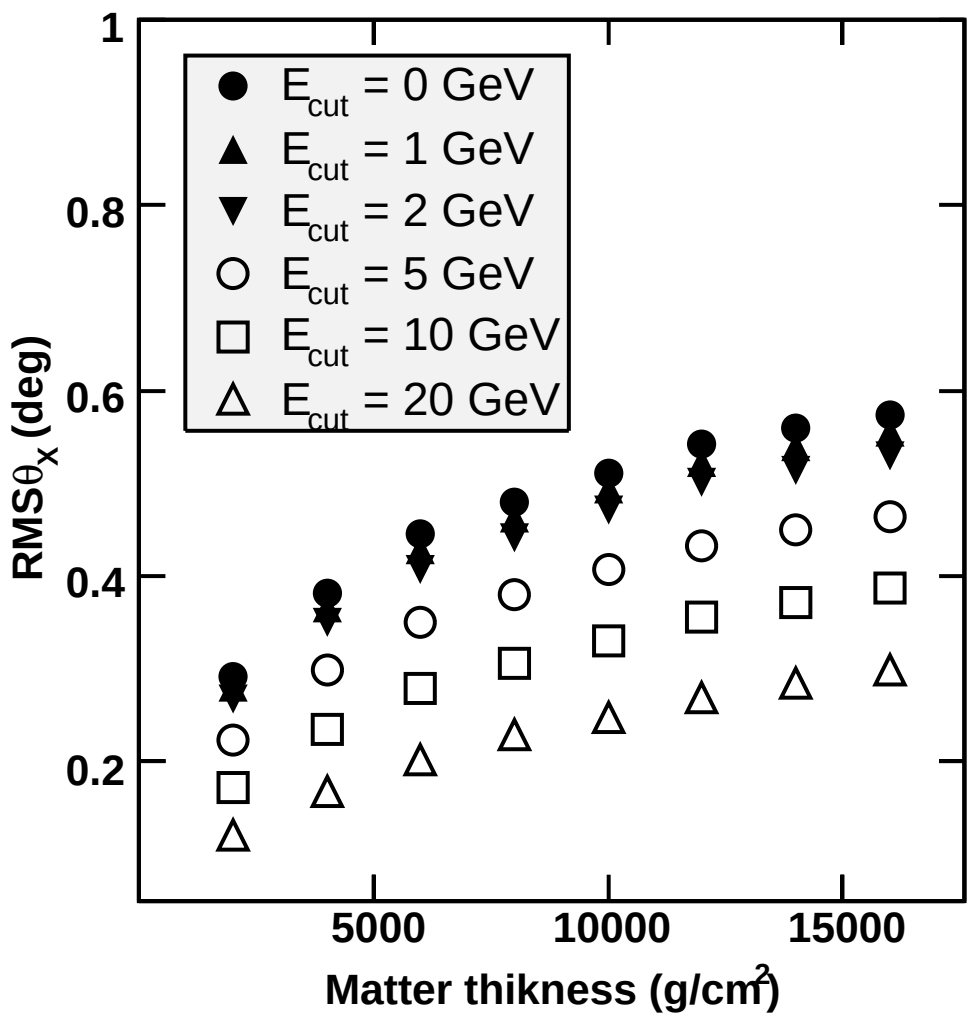

**Fig 5** RMS$_{\theta x}$ dependent on the layer thickness and for different muon energy cuts in underground.

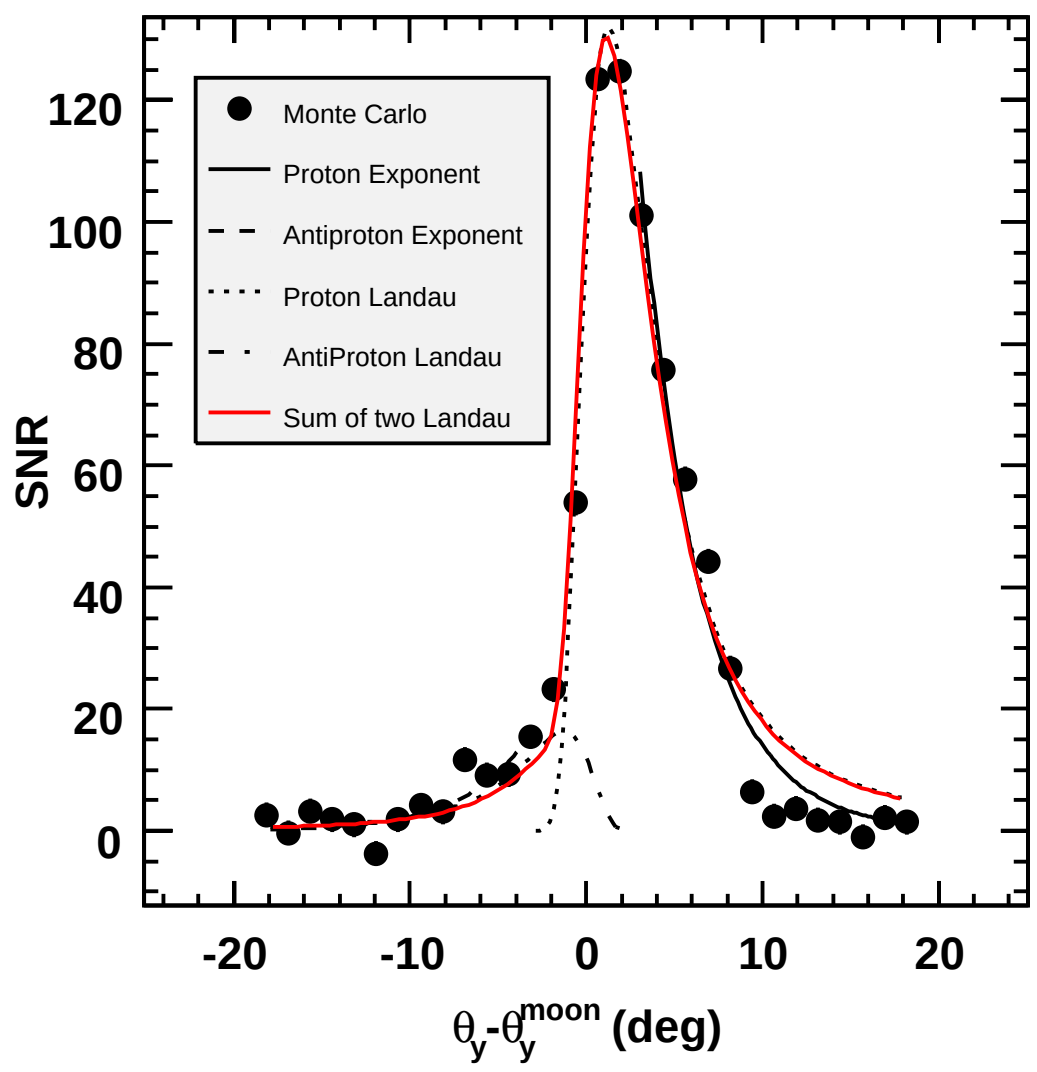

**Fig 6** SNR dependence on $\theta_y$-$\theta_y^{moon}$. Lines are the fit results by the exponential (having the same slope) and Landau (with the same sigma) functions